\documentclass[12pt]{article}
\usepackage{epsfig}
\usepackage{amssymb,graphicx}
\begin{document}

\begin{center}

{\Large \bf {

Comment on 'The operational foundations of PT-symmetric and
quasi-Hermitian quantum theory'

%

 }}

\vspace{13mm}

 {\bf Miloslav Znojil}

 \vspace{3mm}
Nuclear Physics Institute ASCR, Hlavn\'{\i} 130, 250 68 \v{R}e\v{z},
Czech Republic

 e-mail:
  znojil@ujf.cas.cz

%
%
%
%

\end{center}

%

%
\section*{Abstract}

In J. Phys. A: Math. Theor.
55 (2022) 244003,
Alase et al
wrote
that
``the constraint of quasi-Hermiticity on observables''
is not
``sufficient to extend the standard quantum theory''
because
``such a system is equivalent to a standard quantum
system.''
Three addenda
elucidating the current state of the art are
found necessary.
The first one concerns the project:
In the related literature the original ``aim
of extending
standard quantum theory''
has already been
abandoned shortly after its formulation.
The second comment concerns the
method, viz.,
the study in ``the framework of general probabilistic
theories'' (GPT). It is noticed that
a few other,
mathematically consistent
GPT-like theories are available.
The authors do not mention,
in particular, the progress achieved,
under the
quasi-Hermiticity constraint,
in the approach using the effect algebras.
We add that this approach
already found its advanced realistic
applications in the quasi-Hermitian models using the
unbounded operators of observables acting in the
infinite-dimensional Hilbert spaces.
Thirdly, the
``intriguing open question'' about
``what possible constraints, if any, could lead to such a
meaningful extension'' (in the future)
is given an immediate tentative
answer: The possibility is advocated that the desirable constraint
could really be just the
quasi-Hermiticity
of the observables, provided only that
one has in mind its recently developed
non-stationary version.

\newpage

%


%
%
%
%
%

\section{Introduction}


As a part of issue ``Foundational Structures
in Quantum Theory''
the paper ``The operational foundations of PT-symmetric and
quasi-Hermitian quantum theory''
by Abhijeet Alase, Salini Karuvade and Carlo Maria
Scandolo  \cite{Ciftci} fitted very well the scope of
the volume. In a rigorous mathematical style
it offered the readers an interesting material
confirming the
compatibility
between the three recent conceptual
innovations of quantum theory.
Still, we believe that the
authors' coverage of the
subject deserves a few addenda, mainly because
in {\it loc. cit.},
the deeply satisfactory
nature of the
mathematical
analysis seems to be accompanied
by a perceivably less careful
presentation of its implications in
the context of the theoretical quantum physics.

\section{The absence of
extensions of standard quantum theory\label{dve}}

Our first addendum is motivated
by the
last sentence of the abstract in \cite{Ciftci}. It states that
``our results show that neither PT-symmetry nor quasi-Hermiticity
constraints are sufficient to extend standard quantum
theory consistently''. Indeed, it is
rather unfortunate
that this statement
diverts attention from the
very interesting main mathematical
message of the paper
(viz. from the rigorous
confirmation of
compatibility between the three
alternative versions of quantum theory)
to its
much less satisfactory physical contextualization.
The impression
is further strengthened by the last paragraph of the
whole text where we read that
``in conclusion, neither PT-symmetry nor quasi-Hermiticity
of observables leads to an extension of standard quantum mechanics.''
Certainly, non-specialists could
be mislead to interpret such a conclusion
wrongly,
as a disproof of usefulness of
what is usually called PT-symmetric quantum theory
(PTQT, an approach
which is briefly reviewed in section 2.1 of {\it loc. cit.})
or of the so called quasi-Hermitian
quantum theory (QHQT, cf.
its compact review in the subsequent section 2.2 of {\it loc. cit.}).
The misunderstanding seems completed by the
combination of the very first sentence of the abstract with the
very
last sentence of the text:
At the beginning of the Abstract we are told that
``PT-symmetric quantum theory was originally proposed with the aim of extending
standard quantum theory''
(which is not too relevant at present), while the final question reads
``what possible constraints, if any, could lead to such a
meaningful extension'' \cite{Ciftci}.

The main weakness of such a
``theory-extension''
motivation and
of the ``physical'' framing of paper
\cite{Ciftci} is that
the original purpose of ``relaxing the Hermiticity
constraint on Hamiltonians'' (as proposed,
by Bender with Boettcher, in their
enormously influential letter \cite{[1]})
was almost immediately shown overambitious and unfulfilled
(see, e.g., the Mostafazadeh's 2010 very mathematical and detailed
criticism and explanation
``that neither PT-symmetry nor quasi-Hermiticity constraints are
sufficient to extend standard quantum theory'' \cite{[39]}).
Thus,
the authors of \cite{Ciftci} only come with
their ``aim to answer the question of whether
a consistent physical theory with PT-symmetric observables extends standard
quantum theory'' too late. For more than twelve years
the answer is known to be negative \cite{[40]}.

\section{A comment on the method\label{tri}}

Naturally, nobody claims that the PTQT itself is not useful.
Nobody could also deny the relevance and the
novelty of the mathematical results presented in paper \cite{Ciftci}.
It is only a pity
that
its authors
did not better emphasize
how well their analysis
fits
the subject of the special issue,
especially due to their innovative turn of attention
to the so called
general probabilistic theories
(GPT, cf. their compact outline in section 2.3 of \cite{Ciftci}).

Paradoxically, in the GPT context one immediately identifies
the second weakness of the paper. It lies
in a surprisingly short
list of the GPT-approach-representing
references. In the paper the list
just incorporates the eight newer papers~\cite{[27]} - \cite{[34]}
(all of them published after the year 2000)
plus a single older, Foulis-coauthored 1970 paper \cite{[70]}.
Not quite expectedly, the list of references does not contain
any Gudder's results -- after all, paper \cite{Ciftci} is a part of the
special issue which is
explicitly declared to honor
his contribution to the field. Thus, one would expect,
for example, a reference to his later review papers \cite{gudderb,gudder}
where he formulated
one of the key GPT-related
mathematical theses that ``a physical system S under
experimental investigation
and governed by a scientific theory (which may be subject to modification
in the light of new experimental evidence) is represented by a CB-effect algebra''.
An equally unexpected gap in the references also concerns the
absence of the
Foulis' pioneering,
effect-algebras introducing
1994 paper with Bennet \cite{foulisbenn}, or his comparatively recent
review \cite{foulis}. Indeed, both of these papers sought and offered
operational foundations
and gained
insight into the GPT-motivating
relationship between quantum theory
and classical probability theory (this was
emphasized also in \cite{[27]}, etc).

What is an even worse omission is that
the list of references does not contain any other
subject-related
studies like, e.g., paper \cite{paseka}
in which the predecessors of the present authors
considered, {\em explicitly}, the PTQT-GPT
relationship, having reconfirmed that
``from the standpoint of (generalized) effect algebra theory
both representations of our quantum system coincide''.
Similarly, the
QHQT-GPT
relationship may be found studied in paper \cite{riecanova}
in which the
mathematically fairly advanced
analysis incorporated even
the fairly realistic quantum models using unbounded operators.
Indeed, the
rate of the progress is striking, especially when one
recalls just a few years younger report \cite{gudderb}
in which the ``separable complex Hilbert space''
is assumed to be just ``of dimension~1 or more''.

\section{New and promising non-stationary
constraints\label{ctyri}}

At present, it makes sense to accept the fact that
in spite of the robust nature of
the existing
``standard'' formulations of
quantum theory and, in particular, of the quantum mechanics
of unitary systems,
there still exist differences in the
practical applicability of their various
specific implementations. The motivation of the diversity is that
''no [particular]
formulation produces a royal road to quantum mechanics''
\cite{Styer}.
In some sense this implies that
the concept of the ``extension'' of the
existing quantum theory is vague.
The apparently minor technical differences
between the current alternative
formulations of quantum mechanics (as sampled, in \cite{Styer},
on elementary level)
could happen to lead to ``decisive extensions'' in the future.

A good illustrative example can be provided even within the
current stationary forms of
QHQT. Indeed, even in this framework the formalism
can really
be declared equivalent to its standard textbook form.
Still, the equivalence can be confirmed only
under certain fairly detailed and specific mathematical assumptions
(cf.~\cite{[49]}).
These assumptions are, even in the
abstract context of functional analysis, far from trivial
\cite{Dieudonne}.
Paradoxically, even the popular physical quantum
models of Bender and Boettcher \cite{[1]}
have been later found
{\em not\,} to belong to the ``admissible'',
QHQT-compatible class
(see, e.g., \cite{Siegl,Uwe}
for the corresponding subtle details).
Thus, in spite of their manifest and unbroken PT-symmetry,
even these originally proposed benchmark models still wait for a
``meaningful extension'' of their fully consistent
GPT interpretation.

In our third, last addendum we are now prepared to reopen the
vague but important
question of what the words of ``extension''
of the ``standard'' quantum theory could, or do, really mean.
On one side, it is known and widely accepted that
the various existing formulations of
quantum theory
``differ dramatically in mathematical
and conceptual overview, yet each one makes {\em identical\,}
predictions for
all experimental results'' \cite{Styer}.
On the other side, such a rigidity of the theory
is far from satisfactory.
For example, a suitable future amendment of quantum theory
would be necessary
for a still absent
clarification of the concept
of quantum gravity \cite{Rovelli}.

For the sake of brevity
let us
skip here the discussion of the parallel questions concerning the
PT-symmetric quantum models.
This being said we believe that even
the QHQT formalism itself did not say its last word yet.
Indeed,
our optimism
concerning its potential ``theory extension status'' is based on
the recent fundamental clarifications of its scope and structure.
First of all, it became clear that
in the QHQT descriptions of unitary systems it is sufficient
to distinguish just between
their representations in the ``generalized Schr\"{o}dinger picture''
(GSP, stationary and
best presented, by our opinion, in reviews \cite{[49],Carl}
and \cite{[39]})
and in its non-stationary
``non-Hermitian interaction picture'' alternative
(NIP, \cite{NIP,WDW}).
Using this terminology one immediately reveals that
the QHQT-related considerations of paper \cite{Ciftci} just cover the
GSP approach. In other words, the physical inner-product metric
(denoted by symbol $\eta$)
is perceived there as strictly time-independent.
This means that in the GSP language one can easily identify the
(stationary) generator $G$ of the evolution
of the wave functions with the (``observable-energy'') Hamiltonian $H$
(which has real spectrum and is, by assumption, $\eta-$quasi-Hermitian).

The situation becomes different after the
extension of the QHQT
approach to the non-stationary, NIP dynamical regime.
In this case
we will denote the inner-product metric
by another dedicated symbol $\Theta=\Theta(t)$
as introduced in
the first description of NIP in \cite{timedep}.
What is important is that
the observable-energy operator $H=H(t)$
will get split in the sum
of the two auxiliary operators
$G(t)$ and $\Sigma(t)$. As long as they are both neither observable
nor $\Theta$-quasi-Hermitian in general,
we will exclusively assign the name of
the Hamiltonian to the instantaneous energy operator $H$
(with real spectrum),
adding a word of warning
that a different, less consequent terminology
is often used by
some other authors (see, e.g., \cite{Fring,Maamache}).
Even though neither the spectrum of $G(t)$
nor the spectrum of $\Sigma(t)$ is real in general,
the introduction of these operators
endows the NIP formalism
with an additional flexibility,
capable, as we believe, of
opening the new horizons in the contemporary quantum physics:
In the context of relativistic quantum mechanics, for example,
such a hypothetical ``theory-extension''
possibility has been discussed, in detail, in \cite{NIP}.
For the purposes of a potentially new approach to the
problem of the unitary-evolution models of quantum phase transitions
in many-body context,
the formalism has slightly been adapted in \cite{Bishop}.
Last but not least, our very recent paper \cite{WDW}
has been devoted to the possible use of the NIP evolution equations
in a Wheeler-DeWitt-equation-based schematic model of Big Bang in
the context of
quantum gravity and cosmology.
In this spirit, therefore,
certain sufficiently realistic
NIP-based models
could easily happen to acquire an ``extended quantum mechanics'' status,
perhaps,
in the nearest future.

\section{Conclusions}

The key subject discussed in paper \cite{Ciftci} was
the question of
the possible extension of
the scope of quantum theory in general, and
of the realization of such an ambitious
project, in the respective PTQT and QHQT
theoretical frameworks, in particular. In our present commentary
we reminded the readers, marginally, of the existence of
several older, comparably sceptical conclusions as available
in the related literature
(see section \ref{dve} for details).
In section \ref{tri} we then added a few
similar broader-context-emphasizing
remarks
on the mathematical,
GPT-related
aspects of the results of \cite{Ciftci}.
Still, the core of our present message
(as presented in the longest section \ref{ctyri})
concerned physics. We pointed out that at present,
the question of
the possible extension of
the scope of the standard quantum theory
should be considered open
even in the narrower PTQT and QHQT frameworks.

In support of the latter statement we mentioned that

\begin{itemize}

\item
even for the
stationary and,
apparently, most elementary
PTQT potentials
(sampled, say, by the most popular $V(x)={\rm i}x^3$),
the widespread
initial optimism and intuitive ``nothing new''
understanding
of their physical meaning and mathematical background
have both recently been shattered by their
more rigorous mathematical analysis;

\item
one can hardly say ``nothing new''
even in a mathematically much better understood
stationary QHQT {\it alias\,} GSP framework
where, typically, the use of certain stronger assumptions
enables one to circumvent the obstacles
revealed by rigorous mathematics.
Indeed, even in the GSP framework
one can search for an entirely new physics.
Typically, a non-standard phenomenology becomes described by the
QHQT models
in an infinitesimally small vicinity of the so called
exceptional points: Paper \cite{math} offers
an illustrative sample of the
quantum systems which cannot be
described by the standard quantum theory;

\item
in fact, our return to optimism and expectation
that the QHQT may be a ``fundamentally innovative'' theory
found its most explicit formulation in section \ref{ctyri}.
Briefly we exposed there an enormous growth of the
flexibility of the QHQT approach after its ultimate
non-stationary NIP generalization.
In some sense,
the emphasis put upon the deeply promising conceptual nature
of such a flexibility
can be read as the deepest core of our present comment and message.

\end{itemize}

\section*{Data availability statement}

No new data were created or analysed in this study.

\section*{ORCID iD}
https://orcid.org/0000-0001-6076-0093

\end{document}